\documentclass[subscriptcorrection,upint,varvw,barcolor=Goldenrod3,mathalfa=cal=euler,balance,hyphenate,french,pdf-a]{asmejour} %
\usepackage{subcaption}
\usepackage{float}
\usepackage{graphicx}
\usepackage{hyperref}
\usepackage{ulem}
\hypersetup{%
	pdfauthor={Kevin Ma},                       		   	% <=== change to YOUR name[s]!
	pdftitle={LLM for Generating Design Solutions},                  	% <=== change to YOUR pdf file title
	pdfkeywords={LLM Generation, Diversity, Conceptual Design Solutions, Prompt Engineering},% <=== change to YOUR pdf keywords
	pdfsubject = {Generating Design Solutions using LLMs},			% <=== change to YOUR subject
	pdflicenseurl={https://ctan.org/pkg/asmejour},% may delete
}

\JourName{}%<=== change to the name of your journal

\begin{document}

% Change to your author name[s] and addresses, in the desired order of authors.
% First name, middle initial, last name
% Use title case (upper and lower case letters)
% Note usage below for corresponding author.

\SetAuthorBlock{Kevin Ma}{Department of Mechanical Engineering,\\
   University of California, Berkeley,\\
   Berkeley, CA 94720\\
   kevinma1515@berkeley.edu} 

% To label one or more corresponding authors put "Name\CorrespondingAuthor". No space after "Name".
% An optional argument can be added if email is not in address block as
%      "Name\CorrespondingAuthor{write@to.me}"
% Can also include multiple emails and use the command more than once for multiple corresponding authors,
%      "Name\CorrespondingAuthor{write@to.him, write@to.her}"
\SetAuthorBlock{Daniele Grandi}{Autodesk Research\\
   San Francisco, CA, 94111\\
   daniele.grandi@autodesk.com} 

\SetAuthorBlock{Christopher McComb}{Department of Mechanical Engineering,\\
   Carnegie Mellon University,\\
   Pittsburgh, PA 15289\\
   ccm@cmu.edu} 

\SetAuthorBlock{Kosa Goucher-Lambert\CorrespondingAuthor}{Department of Mechanical Engineering,\\
   University of California, Berkeley,\\
   Berkeley, CA 94720\\
   kosa@berkeley.edu} 

%%% Change to your paper title. Can insert line breaks if you wish (otherwise breaks are selected automatically).
\title{Exploring the Capabilities of Large Language Models for Generating Diverse Design Solutions}

%%% Change these to your keywords.  Keywords are automatically printed at the end of the abstract.
%%% This command must come BEFORE the end of the abstract.
%%% If you don't want keywords, omit the \keyword{..} command.

%% Abstract should be no more than 250 words
\begin{abstract}
Access to large amounts of diverse design solutions can support designers during the early stage of the design process. In this paper, we explore the efficacy of large language models (LLM) in producing diverse design solutions, investigating the level of impact that parameter tuning and various prompt engineering techniques can have on the diversity of LLM-generated design solutions. Specifically, LLMs are used to generate a total of 4,000 design solutions across five distinct design topics, eight combinations of parameters, and eight different types of prompt engineering techniques, comparing each combination of parameter and prompt engineering method across four different diversity metrics. LLM-generated solutions are compared against 100 human-crowdsourced solutions in each design topic using the same set of diversity metrics. Results indicate that human-generated solutions consistently have greater diversity scores across all design topics. Using a post hoc logistic regression analysis we investigate whether these differences primarily exist at the semantic level. Results show that there is a divide in some design topics between humans and LLM-generated solutions, while others have no clear divide. Taken together, these results contribute to the understanding of LLMs' capabilities in generating a large volume of diverse design solutions and offer insights for future research that leverages LLMs to generate diverse design solutions for a broad range of design tasks (e.g., inspirational stimuli).
\end{abstract}

\date{2024}%% You can modify this information as desired. 
							%% Putting \date{} will suppress any date.  
							%% If this command is omitted, date defaults to \today
							%% This command must come somewhere before \maketitle

\maketitle %% This command creates the author/title/abstract block. Essential!

%%%%%%%%%%%%%%%%%%%%%%%%%%%%%%%%%%%%%%%%%%%%%%%%%%%%%%%%%%%%%%%%%%%%%%%%%%%%%%%%%%%%%%%%%%%%%%%%%%%%%%%
%%%%%%%%%%%%%%%%%%%%%  End of fields to be completed. Now write! %%%%%%%%%%%%%%%%%%%%%%%%%%%%%%%%%%%%%%

\section{Introduction}

\label{introduction}
Inspirational stimuli have been widely shown to support designers during the early stage design process by serving as a catalyst for creativity and innovation \cite{dahl2002influence, kwon2022enabling, fu2015design}. Among the various methods employed to elicit such stimuli, the use of design examples has proven to be particularly effective \cite{linsey2008modality}. In the past, studies have explored the use of crowdsourcing to retrieve these design examples by leveraging the collective intelligence and diverse perspectives of a large number of individuals to generate a large set of design examples \cite{goucher2019crowdsourcing, yu2014searching}. However, with the recent advances in large language models (LLMs), there has been an increased interest in exploring how LLMs can be used to generate design solutions \cite{ma2023conceptual, zhu2023biologically}. 

Recent advancements in LLMs, such as GPT-4, have opened new avenues for research into their application within the design domain. Through the use of prompt engineering techniques, researchers have demonstrated that LLMs have the capability to produce design solutions that are similar to those conceived from crowdsourced human solutions \cite{ma2023conceptual}. Despite this potential, the solutions generated by LLMs are often less diverse than human-generated solutions, which poses a significant challenge given the importance on novelty and diversity in the context of inspirational stimuli \cite{ma2023conceptual, chan2011benefits}. Therefore, it is essential to identify methods for generating diverse outputs from LLMs if they are to be used as sources of inspirational stimuli. 

\begin{figure*}[h]
\centering
\includegraphics[width=0.9\textwidth]{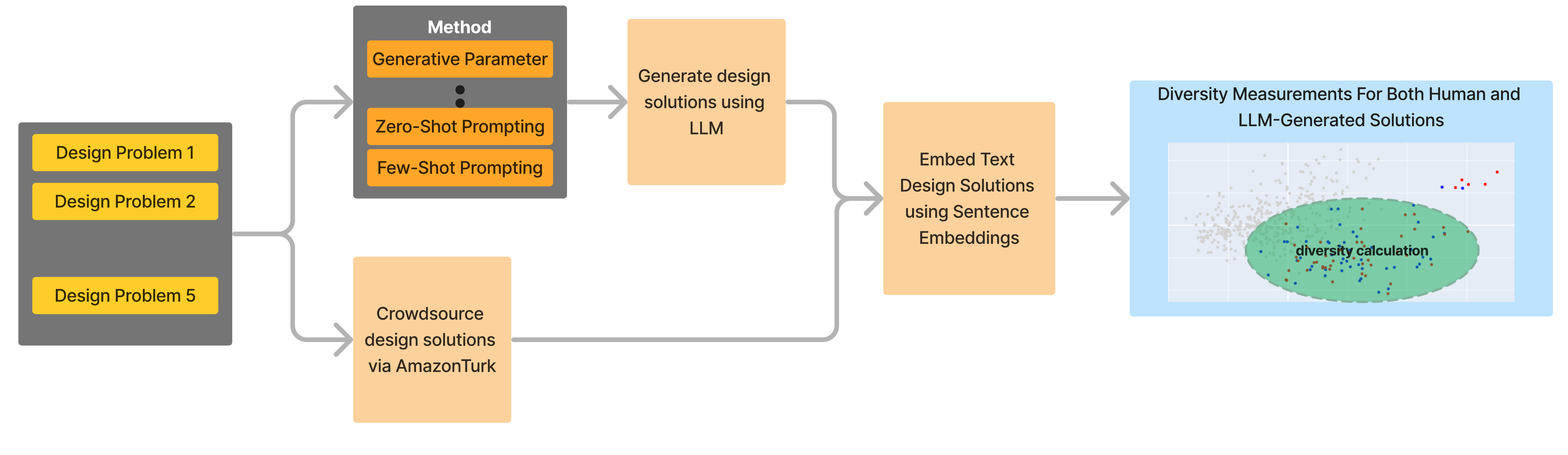}
\caption{\label{fig:figure-1} Our overall objective is to better understand the ability of LLMs to generate diverse design solutions -- tested across a range of design problems and LLM input parameters. For each design topic, we generated 800 total design solutions using a LLM (GPT-4) across different generative parameters (temperature/top-P) and different prompt engineering techniques. Likewise, for each design topic, we retrieved 100 design solutions via crowdsourcing AmazonTurk workers. All the solutions were then converted into vector embeddings, which were used to measure diversity for quantitative comparisons. This was conducted 5 times across 5 different design problems, leading to a total of 4000 design solutions generated by an LLM and 500 design solutions retrieved via Amazon Mechanical Turk crowdsourcing.}
\end{figure*}

Thus, our research was guided by two primary research questions:
\begin{enumerate}
    \item How do parameters that tune LLMs affect the output diversity of the design solutions?
    \item How do different prompt engineering techniques impact the output diversity of the design examples?
\end{enumerate}

In this paper, we explored the use of LLMs to generate a diverse set of design solutions. We first investigated whether systematically varying several parameters within the LLM could affect the diversity of the generated design solution. Then we explored whether or not different prompt engineering techniques could enhance the diversity of the generated design solutions. Our research then involved a comparative analysis of these LLM-generated solutions with those obtained from crowdsourcing platforms, in our case Amazon Mechanical Turk, across five distinct design topics.

We then evaluated the diversity of the generated design solutions through a comprehensive set of computational metrics, comparing various methods of prompt engineering techniques and parameters. Due to initial findings discussed in this paper, our team hypothesized that there may be distinguishable semantic differences between the human and LLM generated design solutions. To investigate this hypothesis, we used logistic regression to analyze whether this was true. The outcomes of this analysis and their implications are discussed.

\section{Background}
\label{background}
% insert something explaining what the background is meant for afterwards...
\subsection{Generating Design Examples for Inspirational Stimuli}
\label{relatedworks-1}
During the early stages of the design process, overcoming design fixation and enhancing creative outcomes is critical. As a result, the generation of design solutions for the purpose of inspirational stimuli has been a focal point of much research in the past \cite{jiang2022data}. In particular, crowdsourcing has emerged as a relevant method for amassing a diverse array of design solutions by drawing upon the collective creativity of a distributed group of individuals \cite{poetz2012value}. Prior research has developed numerous tools to harness the power of crowdsourcing for the retrieval of analogical ideas, with empirical evidence suggesting a positive impact on creativity and ideation \cite{yu2014distributed, yu2014searching}. Notably, crowdsourcing platforms, such as Amazon Mechanical Turk, have demonstrated high capability in quickly generating large arrays of design solutions to be used for inspirational stimuli during the early stages of the design process \cite{goucher2017using}. 

Despite its advantages, crowdsourcing is not without its limitations. A significant challenge lies in the fact that the majority of crowd workers may lack the specialized skills required for design tasks, potentially leading to large amounts of design solutions that are neither useful or feasible \cite{goucher2019crowdsourcing}. However, recent progress in the field of LLMs, especially those utilizing transformer architectures, has shown considerable promise in addressing these concerns. By significantly scaling up LLMs, the quality of responses generated by LLMs has seen marked improvements, which holds the potential to generate higher quality design solutions for inspirational stimuli \cite{brown2020language}. As a result, advances in LLMs has led to a surge of research in exploring the applications of LLMs within the design domain. In particular, our paper utilized GPT-4, a model created by OpenAI that has showcased human-level performances across a variety of professional and academic benchmarks \cite{openai2023gpt4}. Given our use of GPT-4 in this paper, we also acknowledge the rapid pace at which LLMs are advancing, so while our choice of GPT-4 may become outdated in the near future, we argue that our study remains relevant because it aims to investigate the application of LLMs within the design domain. 

Initial studies employing generative transformers, such as GPT-3.5 and GPT-2, have shown that these models are capable of generating design concepts that are both novel and useful \cite{zhu2023generative}. The potential of LLMs also extended to the generation of biologically inspired design concepts, where the language model acted as a bridge, translating biological analogies into natural language to aid in the generation of design concepts \cite{zhu2023biologically}. Furthermore, more recent studies have begun integrating established design frameworks, such as the Function-Behavior-Structure (FBS) frameworks, into LLMs \cite{wang2023task}. A subsequent comparative user study between designers who utilized the FBS format in conjunction with LLMs against those who did not indicated that the integration of FBS with LLMs could be beneficial for ideation \cite{wang2023task}. 

Finally, our recent comparative study between human crowdsourced solutions and LLM-generated solutions revealed that while LLMs are adept at producing solutions that resemble those created by humans, human participants often provided solutions that exhibit greater novelty and diversity \cite{ma2023conceptual}. This serves as a major flaw in LLMs because, in general, we strive for design examples that are diverse in both far-fetched and near-fetched solutions to support designers during the ideation process \cite{goucher2020adaptive}. However, the study in \cite{ma2023conceptual}, which represented an initial examination into the capabilities of LLMs for generating large amounts of design solutions, offered only a preliminary comparison between human and LLM-generated solutions through modest applications of prompt engineering techniques and a constrained exploration of parameter settings. In this paper, we extend on these insights by exploring the impact a larger range of prompt engineering methods and LLM parameters has on the quality of generated solutions. This study seeks to provide a deeper understanding of the differences between LLM-generated solutions and those derived from human crowdsourcing. 

\subsection{Large Language Models}
\label{relatedworks-2}
In this section, we present a short background regarding LLMs, prompt engineering for LLMs, and its associated parameters that can be fine-tuned.

\subsubsection{Pre-Trained Large Language Models}
\label{relatedworks-2.1}
In the realm of natural language processing, LLMs have emerged as a pivotal tool for generating text that is contextually responsive to user input. The advent of transformer-based architectures, coupled with significant advancements in computational power, has facilitated the training of these models on very large datasets, resulting in pre-trained language models capable of emulating human-like reasoning and producing text that closely resembles human writing \cite{brown2020language, vaswani2017attention, wei2022emergent}. 

There exists a plethora of pre-trained language models, with each model exhibiting a distinct range of capabilities in text generation \cite{openai2023gpt4, touvron2023llama, driess2023palm}. However, it is well-known that there is generally a positive correlation between the size of a language model, as measured by the number of its parameters, and the accuracy of its output \cite{brown2020language}. As a result, in this paper, we opted to use GPT-4, a language model developed by OpenAI with a purported model size of over 1 trillion parameters, as the pre-trained language model of choice for our experimental framework. At the time of writing, GPT-4 was the most recent LLM produced by OpenAI. We chose GPT-4 predicated primarily for this reason, as well as the model's popularity and its reported accuracy and performance \cite{openai2023gpt4}. 

\subsubsection{Prompt Engineering and Finetuning Parameters}
\label{relatedworks-2.2}
As a result of the recent advancements from LLMs, a technique known as prompt engineering has emerged. This technique involves the careful manipulation of input prompts to enhance the accuracy and quality of outputs generated by LLMs. One specific technique within this field is few-shot prompting, where examples are provided to the LLMs prior to the generation, which has been shown to improve the accuracy of the generated content \cite{wei2022chain}. Conversely, zero-shot prompting entails querying the language model without prior examples. Research indicates that even minor modifications to zero-shot prompts, such as adding the text \textit{``You are an expert''}, can yield improvements to generated outputs comparable to those achieved through few-shot prompting techniques \cite{kojima2022large}. Our study delved into various prompt engineering strategies examining their effects on the diversity of the generated design solutions. On top of prompt engineering, we also investigated other parameters directly linked to the LLMs, which in our model of choice for GPT-4 would be temperature and top-P. 

Temperature in GPT-4 functions as a control parameter for randomness of generating the predicted texts \cite{openai2023gpt4}. A lower temperature setting results in the model favoring the generation of text with the highest probability, while a higher temperature increases the likelihood of generating less probable text, thus affecting the variability of the generated output. Top-P, on the other hand, determines the breadth of the next chosen consideration of text based on the cumulative probability, ensuring that only the texts with the highest cumulative probability up to a certain threshold are considered \cite{openai2023gpt4}. Thus, a lower Top-P value restricts the model to a narrower selection of highly probable text, whereas a higher value permits a larger selection of less probable text to be considered. In many cases, heuristics are used to tune parameters.

Although these parameters may appear to be highly specific to GPT-4, we argue that studying the impact of these parameters on the generated output diversity is nonetheless crucial. One of the less explored areas in current research surrounding LLMs is the impact that fine-tuning such parameters can have on the LLM's generated output. Our paper contributes to this field by assessing whether adjustments to these parameters significantly influence output generation. We argue that a comprehensive understanding of these effects is essential for designers, particularly when calibrating LLMs to support them during the design process.

\subsection{Measuring Diversity for a Set of Designs}
\label{relatedworks-3}
In our paper, we aimed to quantify the diversity of design solutions generated by both crowdsourcing human workers and LLMs. Quantifying these metrics is important because past research indicated that inspirational stimuli comprising a mix of near and far analogies are most conducive to supporting designers in the early stage design process \cite{fu2013meaning, chan2015impact, cagan2011effective}. Therefore, to assess how broad of a spectrum the generated design solutions cover, we must measure their diversity. 

To computationally measure diversity within a collection of design concepts, we focused on dataset coverage, so concepts that span a wider conceptual space should have higher diversity scores \cite{regenwetter2023beyond}. One straightforward method to evaluate this is by calculating the average distance to the nearest neighbor, also called nearest generated sample \cite{regenwetter2023beyond}. This involves measuring the distance from each original datapoint to its closest generated counterpart and then averaging these distances across the dataset to gauge coverage. Another metric for assessing diversity is the convex hull. This method is predicated on the extent of the spread of the design solutions by using the total hypervolume encapsulated by the convex hull as an indicator of diversity \cite{podani2009convex}. Additionally, determinantal point processes (DPPs) offer an alternative approach. DPPs have been explored in prior research as a suitable metric for evaluating diversity in engineering design and more general machine learning contexts \cite{kulesza2012determinantal, chen2021padgan}. Lastly, we consider the average distance to the centroid of all generated design solutions as an additional potential measure of diversity \cite{regenwetter2023beyond}. In this paper, we applied all these methods to our dataset of design solutions to ensure methodological rigor and consistency across our findings. 

\section{Methods}
\label{methods}
We begin by discussing how we retrieved the human crowdsourced design solutions in Section \ref{methods:crowdsource}. Following this, Section \ref{methods:zero-shot} details the prompt engineering techniques we utilized to generate the design solutions using a LLM. In that same section, we then outlined the various parameter combinations and prompt engineering methods we elected to test. Note, we utilized GPT-4 (version: \textit{gpt-4-0613}) as our choice of LLM for this research study. In Section \ref{methods:diversity}, we describe the diversity metrics selected for analyzing our study. For additional information along with access to the code, we have made the GitHub repository publicly available \footnote{See the source code at \href{https://github.com/kevinma1515/Concept_Generation}{GitHub Repository}}.

\subsection{Crowdsourcing Design Solutions from AmazonTurk Workers}
\label{methods:crowdsource}
All the crowdsourced design solutions were extracted from a previous study as reported by Goucher-Lambert \textit{et al.} where Amazon Mechanical Turk was utilized to solicit design solutions from Amazon Turk workers \cite{goucher2019crowdsourcing}. The goal was to crowdsource a minimum of 100 responses from the workers for each of the following design problems shown in Table \ref{tab:design_problem}.

\begin{table}[h]
\centering
\caption{Design Problems Selected from Historical Data}
\begin{tabular}{l}
\hline
\multicolumn{1}{c}{Design Problems} \\ \hline
1. A lightweight exercise device that can be used while \\ traveling \cite{linsey2014overcoming} \\
2. A device that disperses a light coating of powdered \\ substance over a surface \cite{linsey2008modality} \\
3. A new way to measure the passage of time  \cite{tseng2008role}\\
4. An innovative product to froth milk \cite{toh2014impact}\\
5. A device to fold washcloths, hand towels, and small \\ bath towels \cite{linsey2012design}\\ \hline
\end{tabular}
\label{tab:design_problem}
\end{table}

According to Goucher-Lambert \textit{et al.}, all workers were required to be at least 18 years of age and US citizens. Other than that, no additional demographic information was collected from the workers.

\subsection{Engineering Zero- and Few-Shot Prompts}
\label{methods:zero-shot}
In this section, we outlined the zero- and few-shot prompt inputs into the LLM. In Section \ref{methods:baseline-parameter}, we introduced our approach for baseline zero-shot prompting, which we employed for testing procedures outlined in Sections \ref{methods:baseline-zeroshot} and \ref{methods-fewshot}. Collectively, these sections formed the basis of addressing research question 1 regarding the impact of LLM input parameters on solution diversity. Sections \ref{methods:baseline-zeroshot}, \ref{methods-critique}, and \ref{methods-fewshot} discusses how we used a variety of prompt engineering methods to generate design solutions to address research question 2.

\subsubsection{Baseline Iterative Zero-Shot Prompting \& Parameter Sweeps}
\label{methods:baseline-parameter}
In the baseline zero-shot prompting approach, the initial prompt is formulated as \textit{``Generate 5 design solutions for ''} followed by the specific design problem as outlined in Table \ref{tab:design_problem}. For example, a complete prompt input for a design problem would be \textit{``Generate 5 design solutions for a lightweight exercise device that can be used while traveling''}. Upon receiving the initial prompt input, the LLM generates five design solutions, so both the initial prompt input and the output are recorded in a structured data repository. Subsequently, the LLM receives the prompt input \textit{``Generate 5 more design solutions for''} followed by the same design problem as the initial prompt (e.g., \textit{``Generate 5 more design solutions for a lightweight exercise device that can be used while traveling''}). The responses from this prompt and the prompt input are then added to the data repository. The initial prompt input is conducted just once, but the subsequent prompt inputs are conducted nine total times, resulting in a cumulative total of 50 design solutions for each problem. This iterative prompting strategy was selected to afford the LLM greater latitude in elaborating on each design solutions. 

\begin{figure*}[h]
\centering
\includegraphics[width=0.7\textwidth]{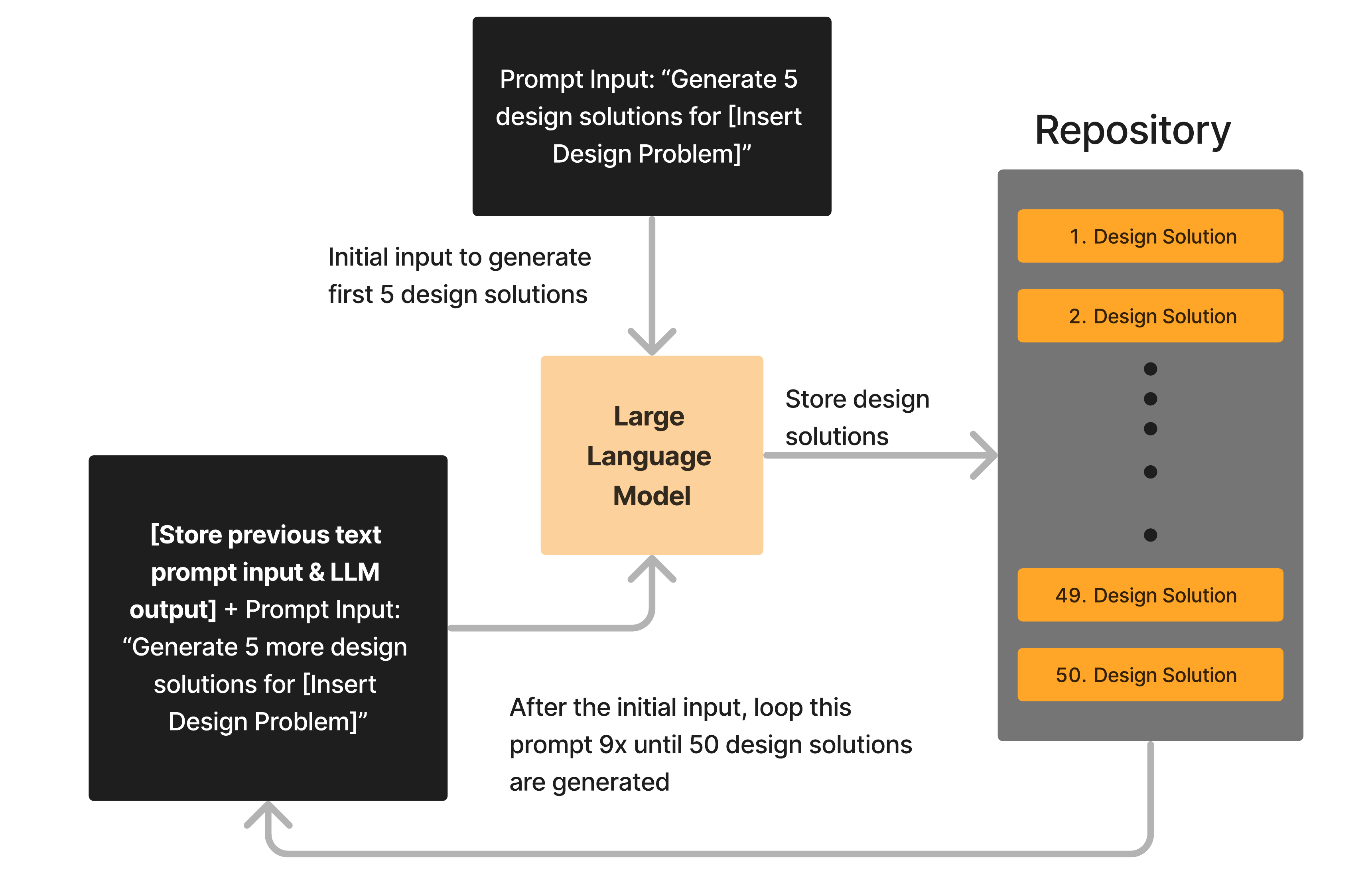}
\caption{\label{fig:zero-shot-baseline} Methodology for zero-shot baseline prompting. To generate a total of 50 design solutions, there is an initial input of \textit{``Generate 5 design solutions for} \textbf{[design problem]}'' (see Table \ref{tab:design_problem} for list of design problems input and Table \ref{tab:zero-shot-prompting} for an example of how the prompts were input). After the LLM (GPT-4 in our case) generates 5 design solutions, they are stored in a data structure. Using the stored data structure, we conditioned the next generation of 5 more design solutions subject to the design solutions already generated as seen in the figure. We performed this loop 9 times until there was a total of 50 design solutions generated.}
\end{figure*}

To address research question 1, we proceeded to use this style of baseline zero-shot prompting for each design problem across several combinations of top-P and temperature parameters. As noted in the GPT-4 API \cite{openai2023gpt4}, the temperature values can vary from 0 to 2 and the top-P values can vary from 0 to 1. In deciding between which parameters to test, we divided the temperature and top-P values to low \textit{(temperature = 0 / top-P = 0)}, medium \textit{(temperature = 1 / top-P = 0.5)}, and high \textit{(temperature = 2 / top-P = 1)}. We then tested the parameters on all possible combinations of low, medium, and high, which is shown in Table \ref{tabular:temp-TopP}.

\begin{table}[h]
\centering
\caption{Temperature and Top-P combination}
\begin{tabular}{c|l}
\hline
Temperature & Top-P \\ \hline
0 & 0 \\ \hline
0 & 0.5 \\ \hline
0 & 1 \\ \hline
1 & 0 \\ \hline
1 & 0.5 \\ \hline
1 & 1 \\ \hline
2 & 0 \\ \hline
2 & 0.5
\end{tabular}
\label{tabular:temp-TopP}
\end{table}

We excluded the high-high combination of temperature and top-P due to the output being incoherent. For each combination, we ran the baseline zero-shot prompting method as discussed in this section. This led to the LLM generating a total of 50 design solutions for each of the 8 different combinations of temperature and top-P.

\subsubsection{Prompt Engineered Baseline Zero-shot Prompting}
\label{methods:baseline-zeroshot}
\begin{table*}[t]
\centering
\caption{Examples of Prompt Engineered Zero- and Few-Shot Prompting for Design Problem 1}
\begin{tabular}{ll}
\hline
Example Prompt & \begin{tabular}[c]{@{}l@{}}Prompt Engineering \\ Type\end{tabular} \\ \hline
\begin{tabular}[c]{@{}l@{}}\textbf{Initial Prompt:} Generate 5 design solutions for a \\ lightweight exercise device that can be used while traveling\\ \textbf{Subsequent Prompts:} Generate 5 more design solutions for a \\ lightweight exercise device that can be used while traveling\end{tabular} & Baseline \\ \hline
\begin{tabular}[c]{@{}l@{}}\textbf{Initial Prompt:} Generate 5 \textcolor{red}{\textbf{novel}} design solutions for a\\ lightweight exercise device that can be used while traveling\\ \textbf{Subsequent Prompts:} Generate 5 more \textcolor{red}{\textbf{novel}} design solutions for a\\ lightweight exercise device that can be used while traveling\end{tabular} & Adjective - Novel \\ \hline
\begin{tabular}[c]{@{}l@{}}\textbf{Initial Prompt:} Generate 5 \textcolor{red}{\textbf{unique}} design solutions for a\\ lightweight exercise device that can be used while traveling\\ \textbf{Subsequent Prompts:} Generate 5 more \textcolor{red}{\textbf{unique}} design solutions for\\ a lightweight exercise device that can be used while traveling\end{tabular} & Adjective - Unique \\ \hline
\begin{tabular}[c]{@{}l@{}}\textbf{Initial Prompt:} Generate 5 \textcolor{red}{\textbf{creative}} design solutions for a\\ lightweight exercise device that can be used while traveling\\ \textbf{Subsequent Prompts:} Generate 5 more \textcolor{red}{\textbf{creative}} design solutions\\ for a lightweight exercise device that can be used while traveling\end{tabular} & Adjective - Creative \\ \hline
\begin{tabular}[c]{@{}l@{}}\textbf{Initial Prompt:} \textcolor{red}{\textbf{You are a design expert.}} Generate 5 design solutions\\ for a lightweight exercise device that can be used while traveling.\\ \textbf{Subsequent Prompts:} \textcolor{red}{\textbf{You are a design expert.}} Generate 5 more \\ design solutions for a lightweight exercise device that can be used\\ while traveling.\end{tabular} & \begin{tabular}[c]{@{}l@{}}Phrase - You are a\\ design expert.\end{tabular} \\ \hline
\begin{tabular}[c]{@{}l@{}}\textbf{Initial Prompt:} \textcolor{red}{\textbf{You are a design expert who is excellent at ideating}}\\ \textcolor{red}{\textbf{far-fetched design ideas.}} Generate 5 design solutions for a \\lightweight exercise device that can be used while traveling.\\ \textbf{Subsequent Prompts:} \textcolor{red}{\textbf{You are a design expert who is excellent at ideating}}\\ \textcolor{red}{\textbf{far-fetched design ideas.}} Generate 5 more design solutions for a \\lightweight exercise device that can be used while traveling.\end{tabular} & \begin{tabular}[c]{@{}l@{}}Phrase - You are a \\ design expert who is\\ excellent at ideating\\ far-fetched design\\ ideas.\end{tabular} \\ \hline
\begin{tabular}[c]{@{}l@{}}\textbf{Initial Prompt:} Generate 5 design solutions for a lightweight exercise\\ device that can be used while traveling. \textcolor{red}{\textbf{Here are some example design}}\\ \textcolor{red}{\textbf{solutions {[}...{]}. Note, the example design solutions are just for guidance.}}\\ \textcolor{red}{\textbf{You do not have to mimic the solutions.}}\\ \textbf{Subsequent Prompts:} Generate 5 more design solutions for a lightweight\\ exercise device that can be used while traveling. \textcolor{red}{\textbf{Here are some example}}\\ \textcolor{red}{\textbf{design solutions {[}...{]}. Note, the example design solutions are just for}} \\ \textcolor{red}{\textbf{guidance. You do not have to mimic the solution.}}\end{tabular} & Few-Shot
\end{tabular}
\label{tab:zero-shot-prompting}
\end{table*}

For this prompt engineering method, we leveraged the zero-shot reasoning capabilities of LLMs \cite{kojima2022large}. Prior studies have demonstrated that prefacing a query with phrases such as \textit{``Let's think step-by-step''} can significantly enhance the LLMs' problem-solving process, yielding results comparable to those achieved through sophisticated prompt engineering methods \cite{kojima2022large}. We sought to determine whether subtle refinements to our baseline zero-shot prompts proposed in Section \ref{methods:baseline-parameter} can stimulate LLMs to produce a more diverse set of solutions. To this end, we experimented with the addition of certain phrases and adjectives to the original prompts. For instance, we introduced statements like \textit{``You are a design expert.''} or \textit{``You are a design expert who is excellent at ideating far-fetched design ideas.''}. Additionally, we incorporated adjectives such as \textit{``novel''}, \textit{``unique''}, \textit{``creative''} within the zero-shot prompts. An example of these prompt modifications is shown in Table \ref{tab:zero-shot-prompting} for design problem 1. We conducted this same process of prompt modification for design problem 1 to all the other design problems listed in Table \ref{tab:design_problem}. 

\subsubsection{Critique Prompt Engineering}
\label{methods-critique}
We borrowed from prior literature that has shown there is benefit in having LLMs critique their initial answers to iterate on the solution and allow the LLM to `reflect' on the answer, provide more rationale behind it, clarify any points of confusion, and add detail to the answer \cite{shinn2023reflexion, white2023prompt}.
From the qualitative evaluations in our prior work, we found that the LLM-generated design solutions tended to lack details compared to the human-crowdsourced design solutions \cite{ma2023conceptual}. This lack of detail made it difficult for experts to evaluate the design solutions for feasibility, novelty, and usefulness. Moreover, design is an iterative process, and we suspected that allowing the LLM to iterate on the solution might result in a more diverse set of solutions.

To implement this critique method we first prompted the LLM to \textit{``Generate 50 design solutions for [design problem]"}, where \textit{[design problem]} is replaced by each of the five design problems in Table~\ref{tab:design_problem}. Then, for each of the 50 design solutions, we prompted the LLM to \textit{``please expand the design solution to add more detail and explain the reasoning and assumptions behind the solution"}. This yielded 50 critiqued design solutions for each of the design problems.

\subsubsection{Few-Shot Prompting}
\label{methods-fewshot}
In past literature, results have shown that the accuracy and quality of the LLMs response can be improved by providing examples within the input prompt prior to requesting a specific output from the LLM \cite{wei2022chain}. This technique is also referred to as  ``few-shot learning''. Our research sought to empirically evaluate this approach for the task of generating a diverse set of design examples, so we employed a similar methodology where we added examples to the initial baseline zero-shot prompt with some modifications (see Table \ref{tab:zero-shot-prompting} under the ``Few-Shot'' category for a specific example). For the selection of the design examples, we opted to randomly sample three design solutions from the human crowdsourced solutions corresponding to its respective design problem. The LLM generated a total of 50 design solutions via this method, and these solutions were then subject to comparative analysis against other prompt engineering strategies discussed in previous sections. 

\subsection{Computationally Measuring Diversity}
\label{methods:diversity}
In this paper, we explored the diversity of design solutions generated through human crowdsourcing and LLM through a variety of prompt engineering techniques. We first converted the textual design solutions into vector embeddings using SentenceBERT \cite{reimers2019sentence}, a model that has been used in past research to capture semantic similarity  \cite{walsh2022semantic, tsumuraya2022topic}. Through this embedding model, each design solution was represented as a 384-dimensional vector, resulting in a 50x384 dimensional embedding space for a set of 50 design solutions. 

Our primary objective was to quantify the diversity within this set of solutions. Because there is no de facto way of measuring diversity, we employed a variety of computational metrics, including determinantal point processes (DPP), nearest generated sample, convex hull volume, and average distance to centroid. These metrics were selected based on established findings in prior research for assessing the spread and coverage of data points in embedding spaces for design related tasks \cite{regenwetter2023beyond}.  While DPP and nearest generated sample calculations can be performed directly on high-dimensional data, other diversity metrics that we used, such as convex hull volume and average distance to centroid, required dimensionality reduction to facilitate computation. As a result, we used principal component analysis (PCA) to dimensionally reduce the embeddings to 20-dimensions for average distance to centroid calculations and to 13-dimensions for convex hull volume calculations. Since we used PCA to reduce the dimensions of our dataset, there is a likelihood that information was lost during this dimensional reduction process \cite{bro2014principal}.

\begin{figure*}[!ht]
\centering
\includegraphics[width=0.98\textwidth]{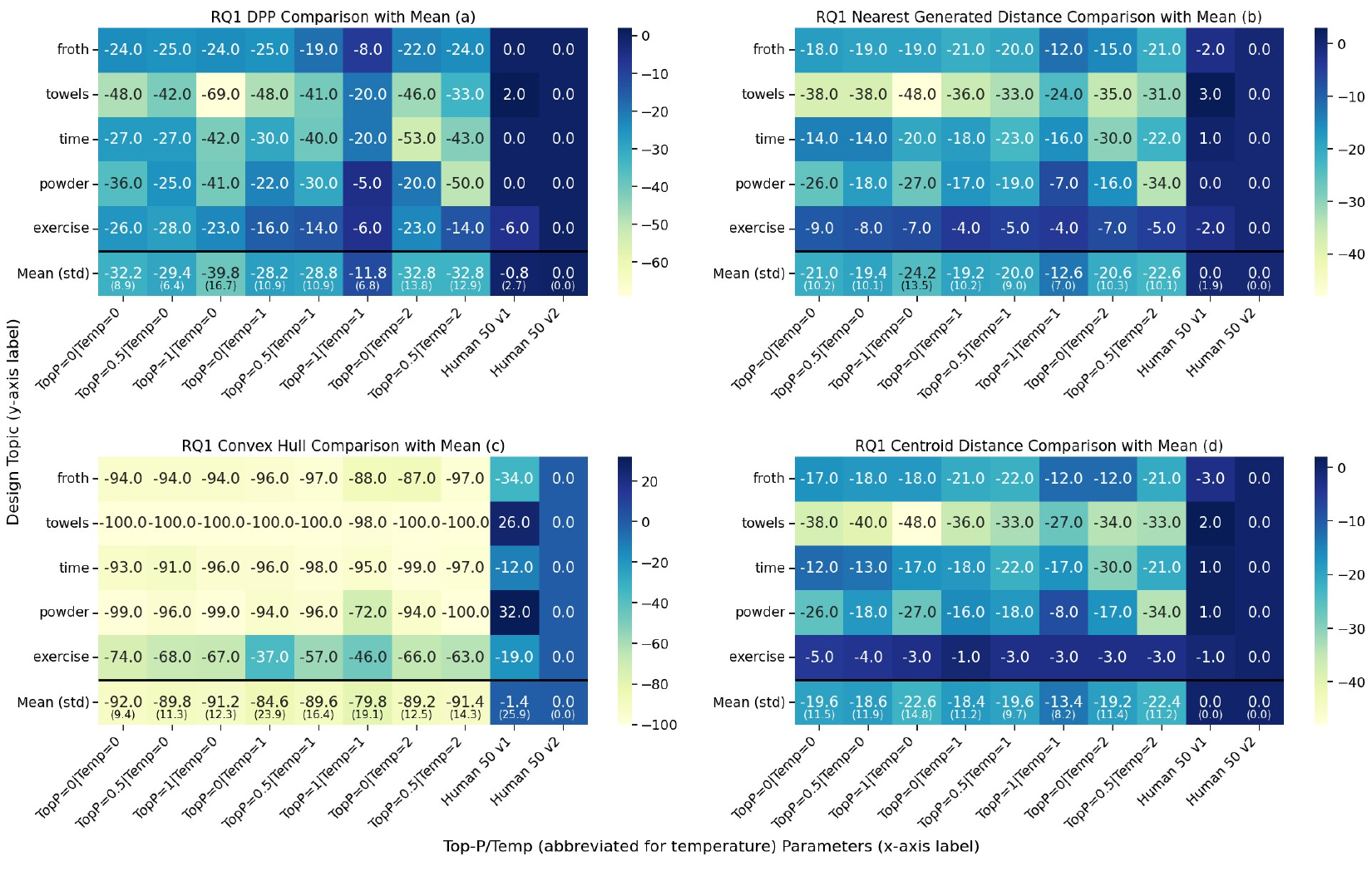}
\caption{\label{fig:results-rq1} There are four total heatmaps, each representing one method of computing diversity. On the x-axis are the temperature and top-P values (see Table \ref{tabular:temp-TopP}, and the y-axis are the corresponding design topics. The tabular value was calculated via percent difference in diversity to `Human 50 v2' measured for the 50 design solutions (see Section \ref{results:parameter-impact} for explanation).}

\includegraphics[width=0.98\textwidth]{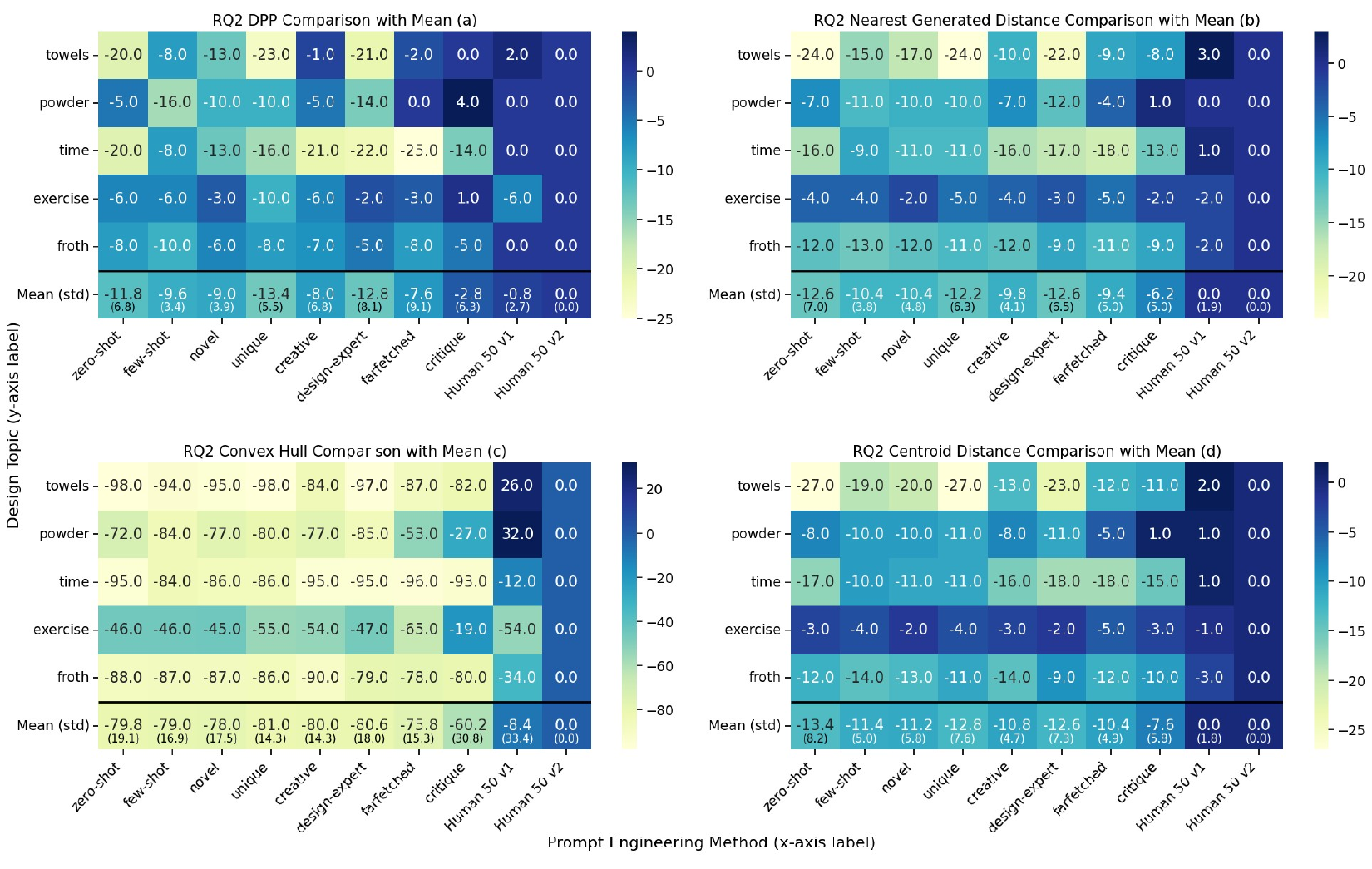}
\caption{\label{fig:results-rq2} There are four total heatmaps, each representing one method of computing diversity. On the x-axis are the different ways of prompt-engineering (see Section \ref{methods}), and the y-axis are the corresponding design topics. The tabular value was calculated via percent difference in diversity to `Human 50 v2' measured for the 50 design solutions (see Section \ref{results:prompt-engineering-impact} for explanation).}
\end{figure*}

In addition, we note that the metrics we employed to measure the diversity of one set of generated design solution often yielded very small values, making them difficult to interpret. To address this, we used percentage change to compare the diversity of each set of parameters for generating design solutions relative to a baseline. Specifically, for each design topic and across all the diversity metrics, we calculated the percentage change for each set of LLM-generated design solutions relative to the second set of human crowdsourced solutions, referred to as `Human 50 v2' in Figures \ref{fig:results-rq1} and \ref{fig:results-rq2}. It is important to note that we divided the 100 total human crowdsourced solutions for each design topic into two groups of 50, which are labelled as `Human 50 v1' and `Human 50 v2' in the heatmaps, respectively. This division was done to ensure a fair comparison between each generated set of LLM-generated design solutions and human crowdsourced solutions. Thus, we used the following equation to calculate the percentage change:

\begin{equation}
\label{equation:percent_change}
\Delta_{x_1 to x_2} = \frac{x_1 - x_2}{|x_2|} \times 100 \%
\end{equation}

where $x_2$ is the score of the second set of the human crowdsourced solution (referred to as `Human 50 v2' in Figure \ref{fig:results-rq1} and \ref{fig:results-rq2}) and $x_1$ is the diversity score for a specific design solution generation method. We opted to use the absolute value of $x_2$ in the denominator because we wanted to avoid instances where the result of the relationship between $x_1$ and $x_2$ is negative due to a negative denominator. 

\clearpage % start a new page 
\clearpage

\section{Results}
\label{results}
% Results are split into two parts: One part is results demonstrating RQ1 and RQ2. Second part is a subsequent exploration on why and how.
Results are split into two parts. Section \ref{results:div} will address the primary research questions introduced in Section \ref{introduction}. As will be discussed in Section \ref{results:div}, we noticed there were differences in the diversity results between the sets of solutions from human crowdsourcing and the ones generated by LLMs. Therefore, Section \ref{results:logistic} will present a follow-up exploration on whether there are semantic embedding differences between the human crowdsourced and LLM generated design solutions.

\subsection{Diversity Results}
\label{results:div}
In the Section \ref{results:interpret}, we will first explain how to interpret the results as presented in Figures \ref{fig:results-rq1} and \ref{fig:results-rq2}. Within that same section, we will also show that the trends in diversity scores are consistent across different methods of computing diversity, which will serve to strengthen some of our findings. For the next two sections, we will address the primary research questions posed in Section \ref{introduction}. In particular, Section \ref{results:parameter-impact} will investigate how parameters that tune LLMs affect the output diversity of the LLM-generated design solutions, while Section \ref{results:prompt-engineering-impact} will impact how different variations of prompt engineering impact the output diversity of the LLM-generated design solutions.  Furthermore, we will highlight and discuss some notable findings in both Figures \ref{fig:results-rq1} and \ref{fig:results-rq2}.

\subsubsection{Interpreting the Results}
\label{results:interpret}
First, we will go over how to interpret the results shown in Figures \ref{fig:results-rq1} and \ref{fig:results-rq2}. Since we used percent difference for our results (see equation \ref{equation:percent_change} and Section \ref{methods:diversity} for explanation), a more negative percent difference signifies worse diversity relative to the second set of human crowdsourced solutions (`Human 50 v2'). Therefore, higher percent differences are indicative of greater diversity based on the metrics used relative to `Human 50 v2'. For example, if we looked at Figure \ref{fig:results-rq1}, specifically within the heatmap titled ``RQ1 DPP Comparison with Mean (a)'', and locating the row corresponding to the design topic ``froth'', the recorded value of ($-24$) means that the 50 LLM-generated design solutions using the parameters TopP = 0 and Temp = 0 has a $-24\%$ difference in DPP score relative to the second set of the human crowdsourced solution, denoted as `Human 50 v2'. This same logic of interpreting the heatmap data would apply across the rest of Figures \ref{fig:results-rq1} and \ref{fig:results-rq2}.

In addition, observations of our results, as shown in Figures \ref{fig:results-rq1} and \ref{fig:results-rq2}, indicated that the trends across the diversity scores remained consistent regardless of the methods used to compute diversity. This uniformity across all four methods suggests that the specific choice of metric does not significantly impact the overall analysis of the diversity assessment. To support this claim, we calculated the Spearman correlation for each design topic across all combinations of the four diversity metric we utilized. The results, as presented in Appendix \ref{append:spearman}, reveals that almost all the results have either strong or moderately strong positive correlations, indicating that the trends observed between different metrics of calculating diversity are closely aligned with each other.

\subsubsection{Parameters Impact on Diversity of Generated Solutions}
\label{results:parameter-impact}
We evaluated the diversity of design solutions generated through the baseline prompt engineering method across eight distinct combinations of temperature and top-P settings (see Section \ref{methods:baseline-parameter}). We then compared these LLM-generated solution to two sets of 50 human crowdsourced solutions (labelled `Human 50 v1' and `Human 50 v2' in Figure \ref{fig:results-rq1} and \ref{fig:results-rq2}), each associated with specific design topics across the y-axis in the heatmaps. 

Findings from Figure \ref{fig:results-rq1} revealed that human crowdsourced solutions (`Human 50 v1' and `Human 50 v2' along the x-axis) had higher diversity scores than LLM-generated solutions regardless of the evaluation method. In addition, the parameter combination of temperature = 1 and top-P = 1 yielded the most diverse set of LLM-generated solutions as evidenced in Figure \ref{fig:results-rq1}. Coincidentally, this parameter combination is also the default configuration provided by OpenAI's GPT-4 in its playground interface. 

In addition, we observed that a definitive trend is not apparent across both the temperature and top-P parameter settings. This observation is supported by Figure \ref{fig:results-rq1}, wherein, upon maintaining a constant temperature and analyzing the variations in diversity score with an increment increase in top-P (across the x-axis), a consistent monotonic trend fails to emerge. Similarly, with top-P held constant, the variation in diversity score with an increase in temperature does not exhibit a clear monotonic pattern either. This observation is of interest, given the general expectation that an increase in temperature should correlate with enhanced ``randomness'' or ``diversity'', and a similar outcome is anticipated with an increase in top-P \cite{openai2023gpt4}. However, the data does not strongly support these assumptions.

\subsubsection{Prompt Engineering Impact on Diversity of Generated Solutions}
\label{results:prompt-engineering-impact}

% remember to talk in the beginning as to why we used temp=1 and top-P = 1 to for our zero-shot comparison
We followed the same procedure as outlined in Section \ref{results:parameter-impact}, but this time we assessed the diversity of design solutions across eight distinct prompt engineering techniques (refer to Table \ref{tab:zero-shot-prompting} for examples of how each prompt engineering method was implemented). Across all prompting engineering methods, we used the temperature and top-P combination from Section \ref{results:parameter-impact} with the highest diversity score, which was determined to be temperature = 1 and top-P = 1. This served as a control variable to evaluate the performance of all the prompt engineering strategies.

Like Section \ref{results:parameter-impact}, we compared the design solutions generated by the LLM to the second set of 50 human crowdsourced solutions, each linked to specific design topics. We then calculated the percentage change in diversity score between the second set of human solutions (denoted as `Human 50 v2' in Figure \ref{fig:results-rq2}) and the design solutions produced by various prompt engineering methods. The results of this comparisons are visually represented in the heatmaps of Figure \ref{fig:results-rq2}.

Similar to findings presented in Section \ref{results:parameter-impact}, the heatmaps in Figure \ref{fig:results-rq2} revealed that human crowdsourced solutions, on average, exhibited higher diversity scores than those generated by the LLMs regardless of the prompt engineering technique applied or the diversity metric used to calculate the score. Notably, the critique-based prompt engineering method yielded the highest average diversity score relative to the other prompt engineering approaches. 

Furthermore, we observed that the diversity metric associated with the LLM-generated design solutions exhibited variability contingent upon the adjectives employed within the prompt inputs. Notably, the adjectives \textit{``creative''}, \textit{``unique''}, and \textit{``novel''}, despite their semantic similarities, have differing outcomes in terms of the diversity scores. Specifically, the use of the word \textit{``unique''} has resulted in comparatively lower diversity scores relative to the words \textit{``creative''} and \textit{``novel''}. Additionally, the inclusion of the phrase \textit{``You are a design expert who is excellent at ideating far-fetched design ideas.''} at the beginning of the baseline prompt led to a considerable improvement in the diversity scores. Moreover, our critique-critique method, which involves adding only one extra prompt sequence that edits the initial 50 generated design solutions, led to the most notable improvement in the diversity scores. These observations suggest that even minor modifications to the prompt's structure or sequence can potentially lead to significant enhancements in diversity scores. These results also hint at the sensitivity of LLMs to subtle adjustments and modifications in the prompts and prompting sequences.

Despite the improvements in the diversity scores of the LLM-generated design solutions, we observed that these solutions did not achieve the level of diversity present in solutions crowdsourced from crowdsourced workers. This observation led us to suspect that there might be semantic differences between design solutions produced by humans and those produced by LLMs. Thus, a subsequent examination of our dataset was undertaken to examine this question, which is detailed in Section \ref{results:logistic}.

\subsection{Logistic Regression Results}
\label{results:logistic}

\begin{table*}[t]
\centering
\caption{Logistic regression model results with confusion matrix and statistical information. True positive and false positive means the model correctly classified LLM-generated solutions or misclassified LLM-generated solutions as human-generated solutions, respectively. Likewise, true negative and false negative means the model correctly classified human-generated solutions or misclassified human-generated solutions as LLM-generated solutions.}
\begin{tabular}{l|ll|r|r|r}
\textbf{\begin{tabular}[c]{@{}l@{}}Confusion\\ Matrix Topics\end{tabular}} & { \textbf{TP/FN}} & {\textbf{FP/TN}} & \multicolumn{1}{l|}{{\textbf{Precision}}} & \multicolumn{1}{l|}{{\textbf{Recall}}} & \multicolumn{1}{l}{{\textbf{F-1 Score}}} \\ \hline
\textbf{Froth}                                                             & 80                   & 0                    & 0.95                                          & 1.00                                       & 0.98                                         \\
                                                                           & 4                    & 16                   & 1.00                                          & 0.80                                       & 0.89                                         \\ \hline
\textbf{\begin{tabular}[c]{@{}l@{}}Exercise\\ Device\end{tabular}}         & 80                   & 0                    & 0.84                                          & 1.00                                       & 0.91                                         \\
                                                                           & 15                   & 5                    & 1.00                                          & 0.25                                       & 0.40                                         \\ \hline
\textbf{Powder}                                                            & 80                   & 0                    & 0.87                                          & 1.00                                       & 0.93                                         \\
                                                                           & 12                   & 8                    & 1.00                                          & 0.40                                       & 0.57                                         \\ \hline
\textbf{Time}                                                              & 80                   & 0                    & 0.89                                          & 1.00                                       & 0.94                                         \\
                                                                           & 10                   & 10                   & 1.00                                          & 0.50                                       & 0.67                                         \\ \hline
\textbf{Towels}                                                            & 80                   & 0                    & 0.91                                          & 1.00                                       & 0.95                                         \\
                                                                           & 8                    & 12                   & 1.00                                          & 0.60                                       & 0.75                                        
\end{tabular}
\label{tab:logistic}
\end{table*}

In the preceding section, we suggested that there may exist semantic differences between LLM-generated design solutions and human-crowdsourced design solutions. To investigate this, we trained a logistic regression model to determine if a decision boundary existed that can distinguish between embeddings of human-crowdsourced and LLM-generated design solutions \cite{hastie2009elements}. 

We selected the prompt engineering dataset from RQ2 for our logistic regression analysis due to the results of the LLM-generated design solutions exhibiting the best diversity value relative to the human-crowdsourced design solutions. To train our model, we first categorized the design solutions as either LLM-generated or human-crowdsourced. Thus, for each design topic, the dataset comprised of 400 LLM-generated design solutions and 100 human-crowdsourced design solutions. We subsequently divided the data for each design topic into training and test sets, allocating $80\%$ for the training set (320 LLM-generated design solutions and 80 human-crowdsourced design solutions) and $20\%$ for the test set (80 LLM-generated design solutions and 20 human-crowdsourced design solutions). A logistic regression model was then trained on the training set for each design topic, and its accuracy was evaluated on the test set. The outcomes, including the confusion matrix, precision, recall, and F-1 score for each design topic, are presented in Table \ref{tab:logistic}. The confusion matrix, as shown in Table \ref{tab:logistic}, can be interpreted as follows: true positive means the LLM-generated solutions are correctly classified, false positive means LLM-generated solutions were incorrectly classified as human-crowdsourced solutions, true negative means the human-crowdsourced solutions were classified correctly, and false negative means the human-crowdsourced solutions were classified incorrectly. 

Interestingly, we observed that all the LLM-generated design solutions were correctly classified, whereas the accuracy in correctly classifying the human-crowdsourced design solutions varied significantly across design topics, as evidenced by the fluctuating recall scores. Given the data imbalance, we argue that the accuracy in classifying human-crowdsourced design solutions served as a more reliable metric for determining the presence of a significant semantic difference between the embeddings of human-crowdsourced and LLM-generated design solutions. As a result, the findings suggests that it is not definitive whether a clear distinction exists between human-crowdsourced and LLM-generated design solutions. The degree of separation appears to vary by design topic, with the milk froth design topic demonstrating the highest accuracy in distinguishing the human-crowdsourced solutions from those generated by LLMs.

\section{Discussion}
\label{discussion}

Our study delved into how adjusting parameters and employing prompt engineering methods influenced the diversity of solutions generated by LLMs, specifically focusing on GPT-4. It was discovered that setting both the temperature and top-P parameters to 1 consistently yielded the most diverse outputs across various diversity metrics and design topics, suggesting the existence of optimal parameter configurations for enhancing output diversity of design solutions for LLMs. Furthermore, we noted that there appeared to be no definitive trend across both the temperature and top-P parameter, defying some expectations about how temperature and top-P should influence the diversity of the LLM-generated design solutions. 

Among the evaluated prompt engineering techniques, the critique-critique approach was found to be the most effective in maximizing solution diversity. Additionally, we observed that slight modifications to adjectives, even when these adjectives are semantically close in meaning, can result in variations in the diversity scores. Furthermore, we noted that the inclusion of brief phrases at the outset of the prompt, aimed at encouraging the LLM to produce more `far-fetched' design solutions, can significantly enhance the diversity score. Moreover, via the critique-critique prompt engineering method, we observed that the introduction of a simple additional prompt layer to revise the initial generated solutions can lead to a greatly enhanced diversity score. These observations suggest that the diversity of the outputs generated by LLMs is possibly sensitive to the structure and wording of the input prompts.

We also observed in our results that regardless of the method we used to improve the diversity of our LLM-generated design solutions, they never exceeded the diversity of human crowdsourced design solutions. This observation led us to further investigate the underlying reasons for this discrepancy. We suspect that semantic differences might exist between the solutions sourced from human participants and those generated by LLMs, but the results are too inconclusive for us to make any significant claims. In this section, we discussed the findings of our results and its potential implications on design. 

\subsection{Enhancing Diversity in LLM-Generated Design Concepts through LLM Parameters}
\label{disc:parameters}
The influence of model parameters on concept design output diversity has not been extensively explored within a formal research context. Various informal sources have posited that parameters such as temperature and top-P in GPT-4 can influence the diversity and creativity of the generated text, which is based on the idea that these parameters can control the likelihood of the subsequent text generation \cite{openai2023gpt4}. Thus, the logic is that the higher the temperature, the more creative or diverse the output. However, findings from our study indicated that this relationship may not be as linear as previously thought. When controlling for one variable and observing changes in temperature or top-P across low, medium, and high settings, our analysis did not reveal a consistent trend in any specific direction. Despite this, our research did identify a parameter combination -- medium temperature and high top-P (temperature = 1 and top-P = 1) -- as having the highest score for diversity. Interestingly, this combination aligns with the default settings provided by OpenAI for GPT-4.

Given the rapid advancements in pre-trained LLMs and the development of new models with potentially different parameter configurations, the question arises: what implications do our findings have? One key takeaway is that our study identified that there are possibly optimal parameters for enhancing diversity in LLM outputs. Other LLMs, such as Gemini, which also have diversity control parameters like temperature, might benefit from similar takeaways when considering how to maximize generated concept solution diversity \cite{team2023gemini}.

However, pinpointing the optimal parameter settings is challenging because our results indicated that there is likely no clear, monotonic trend associated with any specific set of parameters.  We suggest some future approaches that could address this challenge. One approach would involve framing the issue as an optimization problem, where an automatic hyperparameter searching algorithm would aim at maximizing output diversity, possibly through methods like sequential greedy search \cite{bergstra2011algorithms}. Alternatively, we could draw inspiration from prior research that investigated the optimization of machine learning model parameters, including techniques such as reinforcement learning or model-agnostic meta-learning \cite{barto1997reinforcement, finn2017model}, and adapting it to develop algorithms focused on maximizing diversity by adjusting internal parameters related to LLMs. Finally, another possibility could involve creating a new interface mechanism that integrates various LLM parameters specifically to enhance diversity. For this approach to be effective, a precise definition of diversity (e.g., semantic embedding coverage) would be necessary, along with the development of a new parameter designed to control the diversity of the output.

\subsection{Enhancing Diversity in LLM-Generated Design Concepts through Prompt Engineering}
\label{disc:prompt}
The objective of our prompt engineering manipulations was to discern the significance of prompt engineering on the generated diversity. This was grounded on the idea that the construction and manipulation of prompts will play a crucial role here. However, the challenge lied in the vast array of prompt styling and engineering that was available for experimentation. In our study, we explored three types of prompt manipulation.

At the most basic level, we examined the impact modifying similar adjectives (e.g., \textit{``creative''}, \textit{``novelty''}, and \textit{``unique''}) had on the diversity of the generated design solutions. Our findings suggested that there is an impact, albeit varying significant across different design topics. Furthermore, we delved into the influence of phrasing on the diversity of the output. This was tested by incorporating additional phrases such as \textit{``You are a design expert.''} followed by \textit{``You are a  design expert who is excellent at ideating far-fetched design solutions.''}. Prior research had established that LLMs are capable of zero-shot reasoning, and minor adjustments to the prompt can yield comparable accuracy to more complex prompt engineering strategies \cite{kojima2022large}. Our findings align with this but, instead, in the case of diversity, indicating that the modificatios to the phrasing of the zero-shot prompt can significantly affect the diversity of the generated solutions. Lastly, we experimented with a more advanced prompt engineering technique that included a feedback step, prompting the LLM to critique and modify its own generated solutions. This approach yielded the highest diversity score among the methods tested. As a result, we raise the question of what insights these results provide and what the implications might be for future design endeavors utilizing LLMs to achieve diverse outputs.

On one hand, in Section \ref{disc:parameters}, we proposed that language model parameters could be one factor in achieving higher diversity in LLM generated design concepts. However, this approach is notably inefficient and scarcely covered in existing literature. On the other hand, prompt engineering is a more thoroughly investigated field, and future studies could explore how to maximize diversity through more advanced prompt engineering techniques than those discussed in our paper. One recommendation would be to draw inspiration from prior prompt engineering research \cite{white2023prompt, reynolds2021prompt, wei2022chain}, shift its objective function from accuracy to diversity (if the goal of the original method was accuracy), and examine the effects. Another recommendation could be to leverage the conversational nature of LLMs. Our findings suggested that the critique-critique method exhibited the greatest diversity, indicating the potential of LLMs ability to self-reflect and improve upon their outputs. Future research could build on this by treating LLMs as an agent capable of self-conversation to generate diverse outputs or by facilitating a conversation between two LLMs to enhance output diversity. This approach could be informed by grounded design theory, which viewed the ideation and concept generation phase as a conversational process, where designers or agents engage in dialogue with one another, or themselves, to iterate and expand their solution space \cite{schon2017reflective}.

\subsection{Future Directions in Enhancing Diversity using both LLM and Crowdsourcing Data}

In Section \ref{results:logistic}, a subsequent investigation was undertaken because in Sections \ref{results:parameter-impact} and \ref{results:prompt-engineering-impact}, we observed a disparity in the diversity scores between the human-crowdsourced design solutions and LLM-generated design solutions. To explore this possibility, we trained a logistic regression to examine whether there existed a hyperplane that could distinguish between LLM-generated design solutions and human-crowdsourced design solutions. The outcomes of this analysis presented mixed results: for certain design topics, such as milk froth, a hyperplane was more successful in separating the LLM-generated design solutions from those derived through human crowdsourcing, whereas for other subjects, this was not the case. We hypothesize that these variations may be attributed to the specific nature of the design topics. For instance, the concept of a milk frother might be less familiar to some audiences than others, which could lead to a disparity between human-crowdsourced and LLM-generated design solutions. This could be due to the fact that LLMs are likely trained on data that allows it to have a more comprehensive understanding of a milk frother's design and functionality as opposed to the average crowdsource worker on platforms like AmazonTurk. As a whole though, the results from Section \ref{results:logistic} are not definitive, preventing us from drawing any significant conclusions.

However, there are still some interesting perspectives from these results. In the design topics where human crowdsourced solutions are semantically aligned with those generated by LLMs, results suggested that designers could achieve a broad spectrum of diverse solutions simply by experimenting with various prompt engineering techniques. On the other hand, in the cases where there was a clear semantic gap between human and LLM outputs, we see an opportunity for a synergistic approach. By leveraging human crowdsourced solutions as a source of inspirational stimuli, designers could utilize far-fetched solutions from human crowdsourcing in conjunction with LLM-generated solutions to enhance their own conceptual design space. 

The concept of utilizing AI systems in conjunction with human crowdsourcing is not a new idea, however. Recently, there has been a significant amount of research dedicated to exploring how AI can be harnessed to assist designers in co-creation \cite{hwang2022too, kwon2022enabling, inie2023designing}. Likewise, there has also been a substantial amount of research investigating how designers can utilize crowdsourcing as a means of obtaining inspirational stimuli to support designers in concept generation \cite{yu2014distributed, yu2014searching, goucher2019crowdsourcing}. Given this background, a logical progression would be to investigate the synergistic use of AI systems and crowdsourcing to further enhance designer's creativity. This approach was examined in a study by \textit{Kittur et. al.} \cite{kittur2019scaling}, which demonstrated that employing AI to identify and suggest potential analogies from an extensive database, coupled with the involvement of crowdsource workers to evaluate and refine those AI-generated analogies, could significantly increase the number of innovative ideas that can be given to designers as inspirational stimuli. However, in their work, they had to develop their own AI model, with a much smaller training dataset. Whereas now, with the recent advances in technology leading to powerful pre-trained LLMs, we suspect that the integration of the creative capabilities of human crowdsourcing workers with current AI models, such as GPT-4, can lead to an even greater expansion in the generation of innovative ideas.

\section{Limitations}
\label{limitations}
We acknowledge several limitations inherent to our research methodology and scope, and we discuss some potential future works to address these limitations. Firstly, we bring up the fact that our investigation is primarily predicated on the performance of the GPT-4 model, with a specific focus on its temperature and top-P parameters. Thus, the generalizability of our findings to other LLMs, including iterations of future GPT models or non-GPT-based models, is constrained. As a result, our recommendation that the default parameter from GPT-4 model (temperature = 1 and top-P = 1) be used for generating diverse design solutions may not be useful for other LLMs.

Despite this specificity, the research question posed in this paper still contributes to the discussion on the influence LLM parameters have on the generation of diverse sets of design solutions. Our findings indicate that parameter settings do affect the diversity of outputs in the context of GPT-4. And while our study is limited to a single model, it lays the groundwork for subsequent research to investigate whether these effects are consistent across different LLMs. Likewise, another limitation from our paper is that the prompt engineering techniques we employed were tested exclusively on GPT-4, which may limit their applicability to other LLMs. Future research should extend this exploration to a wider array of models to determine the impact of prompt engineering on the diversity of generated solutions more comprehensively.

It is also important to note that LLMs are going to continuously refine and improve over time. Advances in LLMs capabilities may diminish the relevance of the research questions we posed, particularly if future models are developed to inherently produce more diverse design ideas that rival those generated by human crowdsourcing. Nevertheless, our study provides an essential preliminary comparison between LLM-generated solutions and human-crowdsourced solutions, which is critical for the development of future studies in LLM application for engineering design.

\section{Conclusion}
\label{conclusion}
In this study, we explored the impact of tuning parameters in LLMs and the effectiveness of various prompt engineering strategies on the diversity of the generated design solutions. We found that there existed optimal parameters that lead to the highest diversity, and we found that, surprisingly, the relationship between those parameters did not follow a clear pattern. Among the prompt engineering strategies tested, we found that various prompt engineering techniques ranging from modification to the prompt sequence to slight adjective modifications can lead to varying effects on the diversity score, indicating that the diversity of the LLM outputs may be highly responsive to prompt structure, phrasing, and sequencing. We also conducted a subsequent follow-on investigation to test whether human-crowdsourced and LLM-generated design solutions were semantically different, and found mixed results across design topics. Notably, we were able to perfectly classify LLM-generated solutions from the human-crowdsourced solutions. However, the ability to use a linear classifier to correctly classify human crowdsourced solutions from LLM-generated solutions varied greatly based on the specific design topic.

\section*{Acknowledgement}
%% What grants did we use to fund my GSR for Fall 2023? Who do we want to acknowledge?
This work builds upon prior work published at the ASME International Design Engineering Technical Conferences and Computers and Information in Engineering Conference 2023 \cite{ma2023conceptual}. We also thank the members of the Berkeley Institute of Design for their feedback and support.

\appendix
\section{Spearman Correlation}
\label{append:spearman}
We chose Spearman correlation because of its ability to detect monotonic relationships between two datasets, focusing on whether the direction of changes (trends) is similar between two sets of data, without needing to assume a linear relationship between them \cite{spearman1961proof}. In Tables 4 and 5, we analyzed the Spearman correlation for various design topics by measuring the correlation between all the different possible pairwise combinations of methods we used to measure diversity in this paper. Specifically, in the tables, the x-axis represents the range of design topics for which we generated solutions (refer to Table \ref{tab:design_problem} for a list of the topics). The y-axis, on the other hand, displays all possible pairwise combinations of diversity metrics: DPP, Nearest Generated Sample (NGS, as abbreviated in the tables), convex hull, and average distance to centroid (referred to as centroid distance in the tables). We note that a Spearman correlation close to 1 or -1 indicates a strong positive or negative correlation, respectively, while a correlation close to 0 suggests no correlation. The results, presented in Tables \ref{tab:spearman_rq1} and \ref{tab:spearman_rq2} for the temp/top-P study and the prompt engineering study (RQ1 and RQ2), respectively, shows that nearly all results indicate either strong or moderately strong positive correlation, with one outlier indicating weak positive correlations. This indicates that the trends observed with different metrics are closely aligned, reinforcing our argument that the results are consistent across all diversity metrics. 

\clearpage
\begin{table*}[t]
\centering
\caption{Spearman Correlation Score across all Combinations of the Four Diversity Metrics (Temp/Top-P Study)}
\begin{tabular}{l|lllll}
\hline
Combination                                                                 & \begin{tabular}[c]{@{}l@{}}Spearman\\ Correlation\\ Score (froth)\end{tabular} & \begin{tabular}[c]{@{}l@{}}Spearman\\ Correlation\\ Score (towels)\end{tabular} & \begin{tabular}[c]{@{}l@{}}Spearman\\ Correlation\\ Score (time)\end{tabular} & \begin{tabular}[c]{@{}l@{}}Spearman\\ Correlation\\ Score (powder)\end{tabular} & \begin{tabular}[c]{@{}l@{}}Spearman\\ Correlation\\ Score (exercise)\end{tabular} \\ \hline
DPP/NGS                                                                     & 0.770                                                                          & 0.942                                                                           & 0.924                                                                         & 1.000                                                                           & 0.929                                                                             \\ \hline
\begin{tabular}[c]{@{}l@{}}DPP/Convex\\ Hull\end{tabular}                   & 0.633                                                                          & 0.815                                                                           & 0.911                                                                         & 0.988                                                                           & 0.899                                                                             \\ \hline
\begin{tabular}[c]{@{}l@{}}DPP/Centroid\\ Distance\end{tabular}             & 0.654                                                                          & 0.921                                                                           & 0.869                                                                         & 0.982                                                                           & 0.769                                                                             \\ \hline
\begin{tabular}[c]{@{}l@{}}NGS/Convex \\ Hull\end{tabular}                  & 0.951                                                                          & 0.815                                                                           & 0.982                                                                         & 0.988                                                                           & 0.991                                                                             \\ \hline
\begin{tabular}[c]{@{}l@{}}NGS/Centroid \\ Distance\end{tabular}            & 0.960                                                                          & 0.994                                                                           & 0.976                                                                         & 0.982                                                                           & 0.920                                                                             \\ \hline
\begin{tabular}[c]{@{}l@{}}Convex Hull/\\ Centroid \\ Distance\end{tabular} & 0.978                                                                          & 0.815                                                                           & 0.966                                                                         & 0.994                                                                           & 0.934                                                                            
\end{tabular}
\label{tab:spearman_rq1}
\end{table*}

\begin{table*}[t]
\caption{Spearman Correlation Score across all Combinations of the Four Diversity Metrics (Prompt Engineering Study)}
\centering
\begin{tabular}{l|lllll}
\hline
Combination                                                                 & \begin{tabular}[c]{@{}l@{}}Spearman\\ Correlation\\ Score (froth)\end{tabular} & \begin{tabular}[c]{@{}l@{}}Spearman\\ Correlation\\ Score (towels)\end{tabular} & \begin{tabular}[c]{@{}l@{}}Spearman\\ Correlation\\ Score (time)\end{tabular} & \begin{tabular}[c]{@{}l@{}}Spearman\\ Correlation\\ Score (powder)\end{tabular} & \begin{tabular}[c]{@{}l@{}}Spearman\\ Correlation\\ Score (exercise)\end{tabular} \\ \hline
DPP/NGS                                                                     & 0.963                                                                          & 0.978                                                                           & 0.972                                                                         & 0.629                                                                           & 0.836                                                                             \\ \hline
\begin{tabular}[c]{@{}l@{}}DPP/Convex\\ Hull\end{tabular}                   & 0.976                                                                          & 0.910                                                                           & 0.963                                                                         & 0.606                                                                           & 0.619                                                                             \\ \hline
\begin{tabular}[c]{@{}l@{}}DPP/Centroid\\ Distance\end{tabular}             & 0.963                                                                          & 0.913                                                                           & 0.960                                                                         & 0.390                                                                           & 0.776                                                                             \\ \hline
\begin{tabular}[c]{@{}l@{}}NGS/Convex \\ Hull\end{tabular}                  & 0.988                                                                          & 0.945                                                                           & 0.978                                                                         & 0.778                                                                           & 0.873                                                                             \\ \hline
\begin{tabular}[c]{@{}l@{}}NGS/Centroid \\ Distance\end{tabular}            & 1.000                                                                          & 0.950                                                                           & 0.994                                                                         & 0.886                                                                           & 0.960                                                                             \\ \hline
\begin{tabular}[c]{@{}l@{}}Convex Hull/\\ Centroid \\ Distance\end{tabular} & 0.988                                                                          & 0.935                                                                           & 0.966                                                                         & 0.544                                                                           & 0.841                                                                            
\end{tabular}
\label{tab:spearman_rq2}
\end{table*}
\clearpage
%%%%%%%%%%%%%  BIBLIOGRAPHY  %%%%%%%%%%%%%%%%%%%%%%%%%%%%%%%%%%%%%%%%%

%% <=== delete this line - unless you wish to typeset the entire contents of your .bib file.

\bibliographystyle{asmejour}   %% .bst file that follows ASME journal format. Do not change.

\bibliography{asmejour-template} %% <=== change this to name of your bib file

%%%%%%%%%%%%%%%%%%%%%%%%%%%%%%%%%%%%%%%%%%%%%%%%%%%%%%%%%%%%%%%%%%%%%%

%% To omit final list of figures and tables, use the class option [nolists]

\end{document}